\documentclass[
showpacs,
floatfix,
aps,
prl,
twocolumn,
superscriptaddress,
amssymb,
]{revtex4-1}

\makeatletter
\newcommand*{\rom}[1]{\expandafter\@slowromancap\romannumeral #1@}
\makeatother

\usepackage{amssymb}
\usepackage{latexsym}
\usepackage[dvips]{graphicx}
\usepackage{amsmath}
\usepackage{graphicx}
\usepackage{dcolumn}
\usepackage{amsfonts}
\usepackage{bm}
\usepackage{epsfig}

\newcommand{\be}{\begin{equation}}
\newcommand{\ee}{\end{equation}}
\newcommand{\bea}{\begin{eqnarray}}
\newcommand{\eea}{\end{eqnarray}}

\def\pc{\mathcal{P}}

\def\Eac{E}

\def\tst{\tau_\star}
\def\eac{\epsilon}

\def\ttr{\tau}
\def\tem{\tau_{\it em}}

\def\tin{\tau_{\it in}}

\def\tst{\tau_\star}

\def\epseff{\epsilon_{\mathrm{eff}}}

\def\oc{\omega_{\mbox{\scriptsize {c}}}}

\def\tq{\tau_{\mbox{\scriptsize {q}}}}

\def\ttr{\tau}
\def\tem{\tau_{\mbox{\scriptsize {em}}}}

\def\tin{\tau_{\mbox{\scriptsize {in}}}}

\def\ac{\mathcal{A}}

\def\as{S}
\def\aso{S_{\omega}}

\def\rhom{\rho_{\rm MIRO}}
\def\drhom{\delta\rho_{\rm MIRO}}

\def\rhosm{\rho_{\rm sm}}
\def\rhosmo{\rho_{\rm sm\omega}}

\def\drhos{\delta\rho_{\rm SdH}}
\def\drhoso{\delta\rho_{{\rm SdH}\omega}}

\def\a{\alpha}

\def\o{\mathcal{O}}
\def\dt{\mathcal{D}_T}

\def\xt{\mathcal{X}_T}
\newcommand{\req}[1]{Eq.\,(\ref{#1})}

\newcommand{\reqs}[2]{Eqs.\,(\ref{#1}),(\ref{#2})}
\newcommand{\rfig}[1]{Fig.\,\ref{#1}}

\newcommand{\rref}[1]{Ref.\,\onlinecite{#1}}

\newcommand{\rRef}[1]{Reference \,\onlinecite{#1}}

\begin{document}
\title{Shubnikov-de Haas oscillations in 2D electron gas under subterahertz radiation}

\author{Q.~Shi}
\affiliation{School of Physics and Astronomy, University of Minnesota, Minneapolis, Minnesota 55455, USA}
\author{P.~D.~Martin}
\affiliation{School of Physics and Astronomy, University of Minnesota, Minneapolis, Minnesota 55455, USA}
\author{A.~T.~Hatke}
\affiliation{School of Physics and Astronomy, University of Minnesota, Minneapolis, Minnesota 55455, USA}
\affiliation{National High Magnetic Field Laboratory, Tallahassee, Florida 32310, USA}
\author{M.~A.~Zudov}
\affiliation{School of Physics and Astronomy, University of Minnesota, Minneapolis, Minnesota 55455, USA}
\author{J.\,D. Watson}
\affiliation{Department of Physics and Astronomy, Purdue University, West Lafayette, Indiana 47907, USA}
\affiliation{Birck Nanotechnology Center, Purdue University, West Lafayette, Indiana 47907, USA}
\author{G.\,C. Gardner}
\affiliation{Birck Nanotechnology Center, Purdue University, West Lafayette, Indiana 47907, USA}
\affiliation{School of Materials Engineering, Purdue University, West Lafayette, Indiana 47907, USA}
\author{M.\,J. Manfra}
\affiliation{Department of Physics and Astronomy, Purdue University, West Lafayette, Indiana 47907, USA}
\affiliation{Birck Nanotechnology Center, Purdue University, West Lafayette, Indiana 47907, USA}
\affiliation{School of Materials Engineering, Purdue University, West Lafayette, Indiana 47907, USA}
\affiliation{School of Electrical and Computer Engineering, Purdue University, West Lafayette, Indiana 47907, USA}
\author{L.~N.~Pfeiffer}
\affiliation{Department of Electrical Engineering, Princeton University, Princeton, New Jersey 08544, USA}
\author{K.~W.~West}
\affiliation{Department of Electrical Engineering, Princeton University, Princeton, New Jersey 08544, USA}

\begin{abstract}
We report on magnetotransport measurements in a 2D electron gas subject to subterahertz radiation in the regime where Shubnikov-de Haas oscillations (SdHO) and microwave-induced resistance oscillations (MIRO) coexist over a wide magnetic field range, spanning several harmonics of the cyclotron resonance. 
Surprisingly, we find that the SdHO amplitude is modified by the radiation in a non-trivial way owing to the oscillatory correction which has the same period and phase as MIRO.
This finding challenges our current understanding of microwave photoresistance in 2D electron gas, calling for future investigations.
\end{abstract}
\received{October 13, 2014}
\maketitle
\vspace{-0.25in}
When a 2D electron gas (2DEG) is subject to a perpendicular magnetic field $B$ and low temperature $T$, the longitudinal resistivity $\rho$ exhibits Shubnikov-de Haas oscillations (SdHO), owing to a quantum correction
\be
\drhos  = -\as \cos\pi\nu,~\as = 4 \rho_0 \lambda \dt
\,. 
\label{eq.sdho}
\ee
Here, $\nu$ is the filling factor, $\rho_0$ is the resistivity at $B=0$, $\lambda=\exp(-\pi/\oc\tq)$ is the Dingle factor, $\tq$ is the quantum lifetime, $\dt = \xt/\sinh\xt$,  $\xt=2\pi^2 k_B T/\hbar\oc$, $\oc=eB/m^\star$ is the cyclotron frequency, and $m^\star$ is the effective mass.
When a 2DEG is subject to radiation of frequency $\omega = 2\pi f$, $\rho$ also reveals microwave-induced resistance oscillations (MIRO) \citep{zudov:2001a,ye:2001,ryzhii:1970,durst:2003,lei:2003,vavilov:2004,dmitriev:2005,hatke:2009a,konstantinov:2009,konstantinov:2013,zudov:2014} which, according to Refs. \citep{dmitriev:2009b,dmitriev:2012}, are given by 
\be
\drhom  = - 2 \pi \eac \rho_0  \pc \eta \lambda^2 \sin 2\pi\eac\,,
\label{eq.miro}
\ee
where $\eac=\omega/\oc$, $\pc$ is the dimensionless radiation intensity \citep{khodas:2008,khodas:2010,dmitriev:2012,zhang:2014, note:100}, $\eta = \tau/2\tst+2\tin/\tau$, $\tau$ is the transport lifetime, $\tin$ is the inelastic lifetime, and $\tst^{-1}=3\tau_0^{-1}-4\tau_1^{-1}+\tau_2^{-1}$ \citep{note:7}.
When the photoresistance $\drhom$ approaches the dark resistivity $\rho$ by absolute value, the MIRO minima evolve into zero-resistance states \citep{mani:2002,zudov:2003,yang:2003,willett:2004,smet:2005,zudov:2006a,zudov:2006b,konstantinov:2010,dorozhkin:2011}, which are understood in terms of current domains \citep{andreev:2003,auerbach:2005,finkler:2009,dmitriev:2013}.

The majority of MIRO studies have been performed at relatively high $T$ and low $f$, at which SdHO are strongly suppressed.
Extending experiments to higher $f$ \citep{kovalev:2004,du:2004a,smet:2005,dorozhkin:2005b,dorozhkin:2007b,wirthmann:2007,mani:2007,tung:2009,kvon:2014} and lower $T$ yields a regime where SdHO and MIRO coexist, allowing to explore possible mixing between these two types of quantum oscillations and to investigate the effect of radiation on SdHO in general.

It has been known for some time that microwaves suppress SdHO in the vicinity of the cyclotron resonance, $\eac \approx 1$ \citep{kovalev:2004,dorozhkin:2007b,du:2004a}.
As SdHO are sensitive to the thermal smearing of the Fermi surface, this suppression can be directly linked to absorption, which is indeed the strongest near the cyclotron resonance \citep{dmitriev:2003,lei:2005a,lei:2005b}.
Away from the cyclotron resonance, our understanding of how microwaves affect SdHO is definitely lacking.
Some experiments have shown that the effect of microwaves on the SdHO is the weakest near half-integer $\eac$,
which was attributed to the suppression of both inter- and intra-Landau level absorption \citep{dorozhkin:2005b,dorozhkin:2007b}.
Another experiment \citep{mani:2007} found that as the MIRO minima approach zero, the SdHO amplitude vanishes in proportion with the background resistance.
\rRef{mani:2007} then argued that in an irradiated 2DEG, $\rho_0$ in \req{eq.sdho} should be replaced by $\rhom \approx \rho_0 + \drhom$.

There exist several mechanisms that could lead to modification of the SdHO by radiation. 
First, the absorption coefficient $\ac$ is expected to acquire an oscillatory quantum correction \cite{ando:1975a,abstreiter:1976,dmitriev:2003,vavilov:2004,raichev:2008,fedorych:2010} which, according to \rref{dmitriev:2003,fedorych:2010}, is given by
\be
\delta \ac_{\rm q} \simeq 2\ac_D \lambda^2\cos2\pi\eac\,,
\label{eq.ma}
\ee
where $\ac_D$ is a classical absorption described by a Drude formula \cite{note:300,fedorych:2010,zhang:2014}.
Since oscillations in $\ac$ translate to oscillations in $T$ \citep{lei:2005a,lei:2005b}, \req{eq.ma} suggests that the microwave-induced suppression of SdHO is maximized near the cyclotron resonance and, to a much lesser extent, near its harmonics,
\be
\eac = n = 1,2,3,...\,.
\label{sdhac1}
\ee

%Oscillations in electron T \citep{lei:2005a,lei:2005b}
In addition, theory also predicts a radiation-induced oscillatory correction, of the order $\o(\lambda)$ \citep{note:1}, to the dc resistivity.
While the inelastic mechanism produces no such contribution \citep{dmitriev:2011}, the displacement mechanism dictates that $\as$ in \req{eq.sdho} acquires an oscillatory correction and should be replaced by \citep{lei:2009b,dmitriev:2011}
\be
\aso  =  \as \left [ 1 -  \pc \frac \ttr \tst \sin^2 (\pi \eac) \right ]\,,
\label{eq.dis}
\ee 
suggesting that the SdHO amplitude is minimized at
\be
\eac = n+1/2 = 3/2,5/2,7/2,...\,,
\label{sdhac2}
\ee
a condition orthogonal to \req{sdhac1}.
Finally, the same condition, \req{sdhac2}, can be expected from classical oscillations in magnetoabsorption \cite{dmitriev:2004}, $\delta\ac_{\rm c}/\ac_D \sim -\cos 2\pi\eac$, which can be stronger than quantum oscillations, given by \req{eq.ma}, in a typical 2DEG. 

In this Rapid Communication we experimentally investigate the photoresistance in high-quality 2DEGs.
Using high $f$, low $\pc$, and low $T$ allows us to overlap MIRO and SdHO over a wide range of $\eac$ and to investigate the SdHO waveform near multiple harmonics of the cyclotron resonance.
Our data reveal pronounced modulation of the SdHO amplitude which persists to $\eac \approx 6$.
Surprisingly, even though the modulation is periodic in $\eac$, it cannot be described by
 either \reqs{eq.ma}{sdhac1} or \reqs{eq.dis}{sdhac2}.
Instead, the radiation-modified SdHO amplitude closely replicates the MIRO waveform, see \req{eq.miro}, suggesting a non-trivial mixing of MIRO and SdHO.
While it is well established that quantum oscillations of the order $\o(\lambda^2)$ interfere with each other \citep{zhang:2007c,hatke:2008a,hatke:2008b,zhang:2008,khodas:2008,khodas:2010,wiedmann:2010b,raichev:2010b,dmitriev:2010}, the observed correlation between MIRO$\sim\o(\lambda^2)$ and SdHO$\sim \o(\lambda)$ is totally unexpected \citep{note:202}.

%intro, previous studies
While we have obtained similar findings from a variety of samples grown at Princeton and Purdue, in what follows we present the results from two Purdue-grown Hall bars, \rom{1} and \rom{2}, of width $w_{\rm \rom{1}} = 300$ $\mu$m and $w_{\rm \rom{2}} = 200$ $\mu$m, respectively. 
Sample \rom{1} is fabricated from a 30 nm-wide Al$_{0.0015}$Ga$_{0.9985}$As/Al$_{0.24}$Ga$_{0.76}$As quantum well, with density $n_e \approx 3.1 \times10^{11}$ cm$^{-2}$ and mobility $\mu\approx 3.6 \times10^{6}$ cm$^2$/Vs.
Sample \rom{2} contains a 30 nm-wide GaAs/Al$_{0.24}$Ga$_{0.76}$As quantum well, with $n_e \approx 2.6 \times10^{11}$ cm$^{-2}$ and $\mu \approx  2.1 \times 10^{7}$ cm$^2$/Vs.
The resistivity was measured using a standard low-frequency lock-in technique, in sweeping $B$, with $f$ from 0.2 to 0.4 THz, generated by backward wave oscillators.

To facilitate the discussion of our results, we first define the relevant quantities and introduce convenient notations.
In the absence of microwave radiation, the resistivity can be represented as
\be
\rho = \rhosm + \drhos\,,
\label{dark}
\ee
where $\rhosm$ is the smooth part of the resistivity \citep{note:3} and $\drhos$ is given by \req{eq.sdho}.
When the radiation is present, we write the resistivity as 
\be
\rho_{\omega} =  \rhom + \drhoso\,, 
\label{rhoomega}
\ee 
where we have introduced $\rhom = \rhosmo + \drhom$ containing slowly varying background $\rhosmo$ \citep{note:5} and MIRO photoresistance, see \req{eq.miro}.
The main goal of our study is to examine if and how $\drhoso$ is different from $\drhos$.

%%%%%%%%%%%%%%%%%%%%%%%%%%%%%%%%%%%
\begin{figure}[t]
\includegraphics{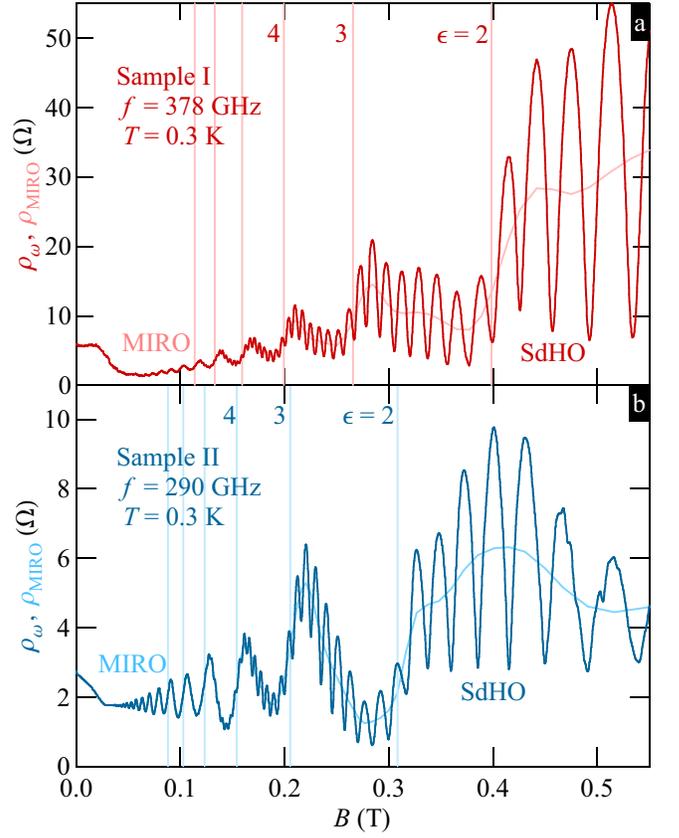}
\vspace{-0.1 in}
\caption{(Color online) 
(a) [(b)] Magnetoresistivity $\rho_\omega(B)$ (dark curves) measured in sample \rom{1} [B] irradiated by microwaves of $f = 378$ GHz [$f = 290$ GHz] at $T = 0.3$ K. 
Vertical lines are drawn at the cyclotron resonance harmonics, $\eac = \omega/\omega_c = 2,3,4,...\,$.
Both panels also show $\rhom(B)$ (light curves) obtained by averaging out SdHO, see \req{rhoomega}.
}
\vspace{-0.15 in}
\label{fig1}
\end{figure}
%%%%%%%%%%%%%%%%%%%%%%%%%%%%%%%%%%%

In \rfig{fig1}(a) [(b)] we present the magnetoresistivity $\rho_\omega(B)$ measured in sample \rom{1} [\rom{2}] irradiated by microwaves of $f = 378$ [290] GHz at $T = 0.3$ K. 
The vertical lines are drawn at the cyclotron resonance harmonics, $\eac = \omega/\omega_c = 2,3,4,...\,$. 
In both samples, the data reveal pronounced MIRO, persisting down to $B \approx 0.05$ T, and SdHO, terminating around $B \approx 0.15$ T.
Owing to high $f$ and low $T$, there exists a wide range of $B$ where SdHO and MIRO coexist.
Most importantly, this range extends over several harmonics of the cyclotron resonance, spanning up to $\eac \approx 6$ and $\eac \approx 5$ for sample \rom{1} and B, respectively.
We immediately notice that, under irradiation, the SdHO amplitude $\aso$ is {\em not} a monotonic function of $B$, in contrast to the ``dark'' amplitude $\as$ described by \req{eq.sdho}.
We thus conclude that the effect of radiation on SdHO is not limited to non-resonant heating.

%%%%%%%%%%%%%%%%%%%%%%%%%%%%%%%%%%%
\begin{figure}[t]
\includegraphics{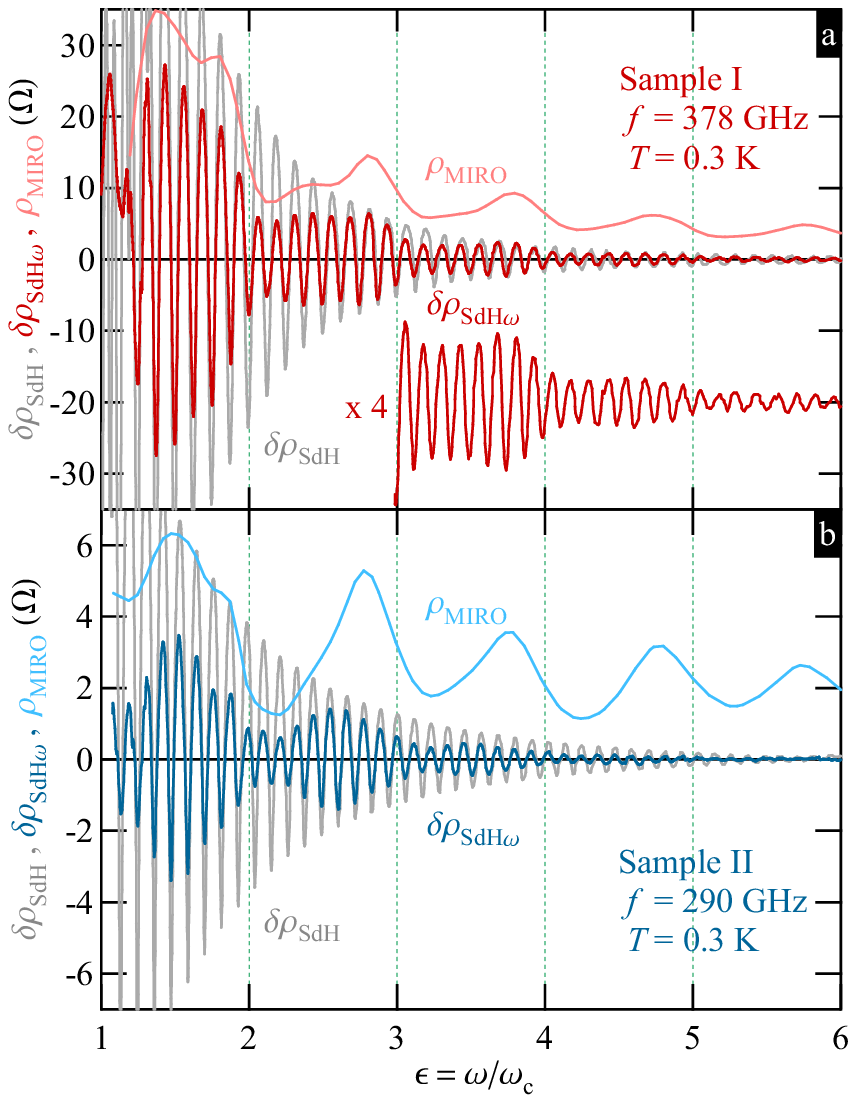}
\vspace{-0.1 in}
\caption{(Color online) 
(a) [(b)] $\drhos$, $\drhoso$, and $\rhom$, obtained from the data in \rfig{fig1}(a) [\rfig{fig1}(b)], versus $\eac$.  
Panel (a) also shows $\drhoso$, multiplied by 4 and offset by 20 $\Omega$.
 }
\vspace{-0.2 in}
\label{fig2}
\end{figure}
%%%%%%%%%%%%%%%%%%%%%%%%%%%%%%%%%%%
To get more insight into the radiation-induced changes of SdHO, one needs to separate $\rhom$ and $\drhoso$, entering \req{rhoomega}.
Since $\rhom$ oscillate much slower than SdHO [cf. \req{dark}], it can be easily obtained by averaging out faster SdHO.
Obtained in this way, $\rhom(B)$ is shown in both panels of \rfig{fig1} by light curves running midway between the SdHO maxima and minima.

Having found $\rhom$, we now use \req{rhoomega} to obtain $\drhoso$ by subtracting $\rhom$ from $\rho_\omega$, both shown in \rfig{fig1}. 
The results for sample \rom{1} and \rom{2} are presented as a function of $\eac$ in \rfig{fig2}(a) and (b), respectively.
For comparison, we also include $\rhom$ and $\drhos$, as marked.
The latter was found using \req{dark} by subtracting the smooth part of the resistivity $\rhosm$ from $\rho(B)$ measured without irradiation.
Direct examination of the SdHO reveals that $|\delta\rho_{{\rm SdH}\omega}| \leq |\delta\rho_{\rm SdH}|$ in the entire range of $\eac$, which is expected because radiation elevates the temperature.
In addition, one can now clearly see that, in contrast to $\drhos$, which monotonically decays with $\eac$ in accordance with \req{eq.sdho}, $\drhoso$ exhibits clear signs of modulation with the period close to unity.
Furthermore, a comparison of $\drhoso$ and $\rhom$ hints on strong correlation between the two quantities.

%%%%%%%%%%%%%%%%%%%%%%%%%%%%%%%%%%%
\begin{figure}[t]
\includegraphics{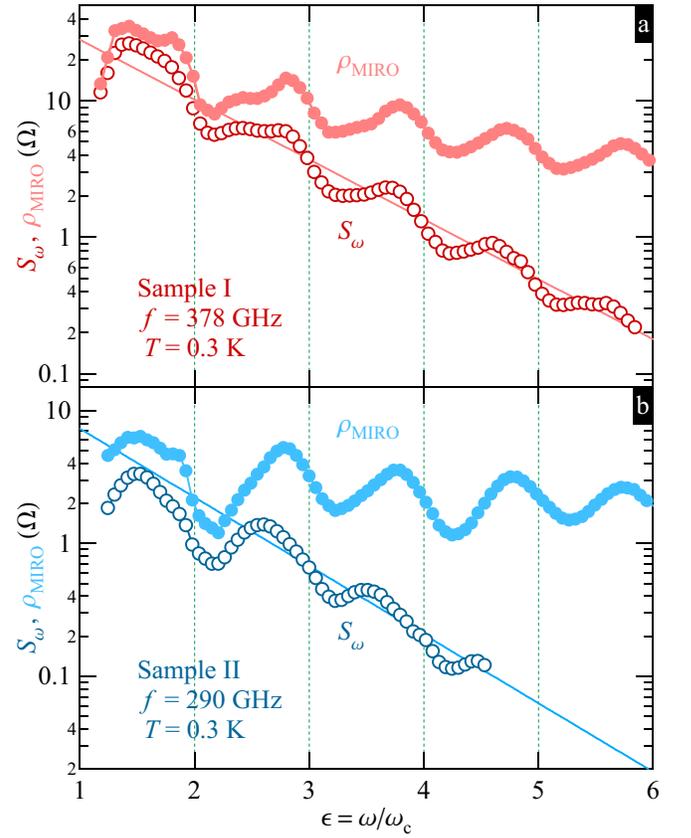}
\vspace{-0.1 in}
\caption{(Color online) 
(a) [(b)] $\rhom$ (solid circles) and $\aso$ (open circles), extracted from the data in \rfig{fig1}(a) [\rfig{fig1}(b)], as a function of $\eac$.
}
\vspace{-0.15 in}
\label{fig3}
\end{figure}
%%%%%%%%%%%%%%%%%%%%%%%%%%%%%%%%%%%

We next extract the amplitude of $\drhoso$, shown in \rfig{fig2}, and examine it in more detail.
In \rfig{fig3} we present the extracted amplitude $\aso$ (open circles) and $\rhom$ (solid circles) as a function of $\eac$ on a log-linear scale.
Once plotted together, the correlation between $\aso$ and $\rhom$ becomes very clear $-$ both quantities oscillate in phase with each other.
In other words, radiation induces minima in SdHO amplitude at
\be
\eac \approx n+1/4 = 5/4,9/4,13/4,...\,,
\label{sdho3}
\ee
in contrast to both the scenario considering oscillations in magnetoabsorption [\req{eq.ma}] and the one
predicting direct modification of the SdHO [\req{eq.dis}].

%%%%%%%%%%%%%%%%%%%%%%%%%%%%%%%%%%%
\begin{figure}[t]
\includegraphics{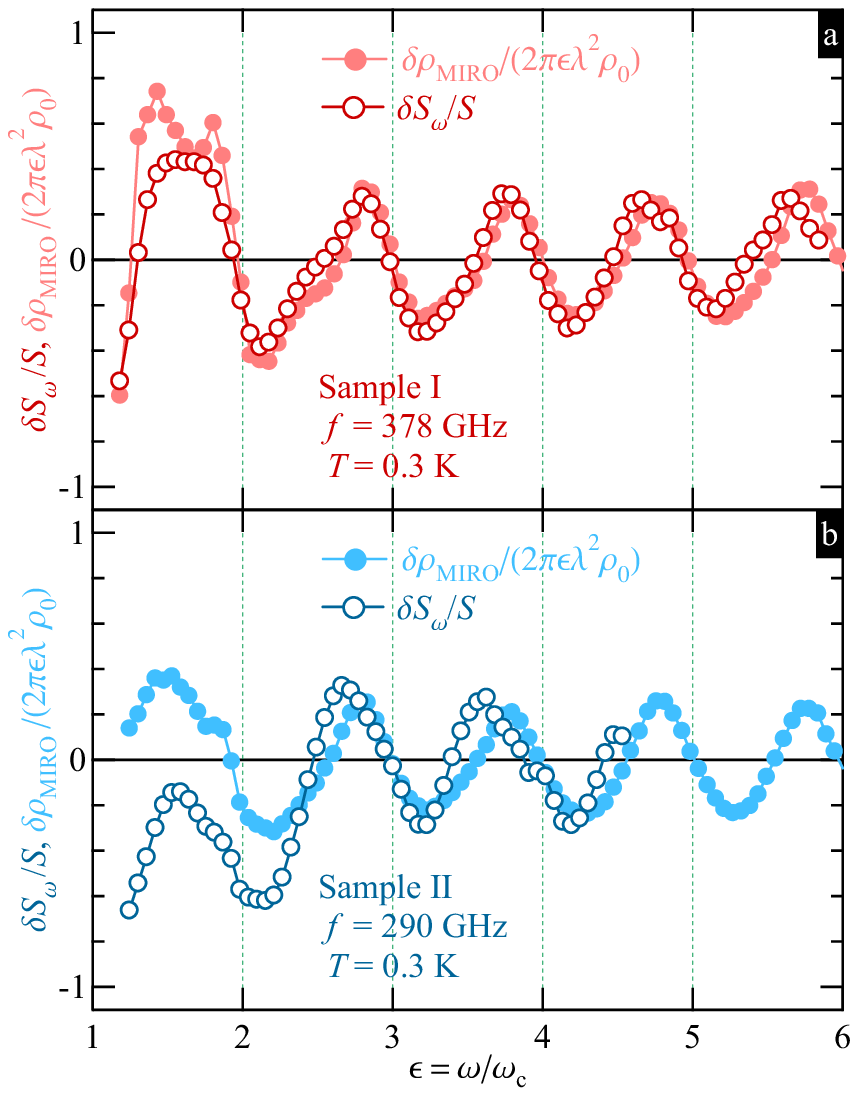}
\vspace{-0.1 in}
\caption{(Color online) 
(a) [(b)] $\delta \aso/\as$ (open circles) and $\drhom/(2\pi \eac \lambda^2 \rho_0)$ (solid circles), extracted from the data in \rfig{fig1}(a) [\rfig{fig1}(b)], as a function of $\eac$. }
\vspace{-0.15 in}
\label{fig4}
\end{figure}
%%%%%%%%%%%%%%%%%%%%%%%%%%%%%%%%%%%

In the remaining part of this Rapid Communication we search for an empirical relation describing the SdHO amplitude in the presence of radiation.
To this end, we extract and compare the oscillatory parts in $\aso$ and in $\rhom$.
More specifically, we introduce the dimensionless quantity $\delta \aso /\as = \aso/\as - 1$, where $\as$ is the smooth, non-oscillating part of the SdHO amplitude shown in \rfig{fig3} by straight lines, and present the results in \rfig{fig4}. 
For comparison, we also plot the oscillatory part of MIRO, $\drhom/(2\pi \eac \lambda^2 \rho_0)$.

We immediately see that both quantities oscillate around zero without noticeable decay, confirming that exponential factors have been properly eliminated.
As already anticipated, a very good agreement in both the period and the phase is found for the whole range of $\eac$ studied.
This finding indicates that observed oscillations in SdHO amplitude are closely related to MIRO and thus are unlikely to originate from resonant heating caused by oscillations in magnetoabsorption [\req{eq.ma}].  
We thus conclude that under presence of radiation the SdHO amplitude is given by
\be
\aso \approx \as (1- \a \sin 2 \pi\eac)\,,
\label{eq.s}
\ee
where $\a$ is a dimensionless $\epsilon$-independent constant, showing that $\delta\aso$ is a correction of order $\mathcal{O}(\lambda)$, just like SdHO themselves. 

The most surprising finding, however, is a \emph{quantitative} correlation between $\delta \aso$ and $\drhom$, namely 
\be
\frac {\delta \aso}{\as} \approx \frac{\drhom}{2\pi \eac \lambda^2 \rho_0} \approx - \alpha \sin 2 \pi \eac\,,
\label{eq.alpha}
\ee
where all parameters have been obtained experimentally.
The observed correlation is completely unexpected and is found almost everywhere, except at $\eac \lesssim 2$ in sample \rom{2}. 
The latter can be linked to increased absorption close to the cyclotron resonance, where SdHO are suppressed due to resonant heating \citep{kovalev:2004,dorozhkin:2007b,du:2004a,note:301}.
The absence of such deviation in sample \rom{1} can be attributed to considerably higher $f$ which reduces the influence of the cyclotron absorption peak.
Interestingly, combining \req{eq.alpha} with \req{eq.miro} one finds that $\alpha \approx \pc \eta$.

Finally, we examine the proposal of \rref{mani:2007} that the SdHO under irradiation can be described by \req{eq.sdho} with $\rho_0$ replaced by $\rho_0 + \drhom$ \citep{note:101}. 
Taking this approach, one obtains $\delta \aso/\as = \drhom/\rho_0$, a result similar to \req{eq.alpha}, but with an extra factor $2\pi\eac\lambda^2$, which has significant dependence on $\eac$. 
Indeed, as $\eac$ increases from 2 to 6, $2\pi\eac\lambda^2$ decreases by nearly a factor of 5 for sample \rom{1}.  
In contrast, our data shown in \rfig{fig4}(a) show virtually no decay at $\eac \gtrsim 2$.
In addition, if this factor were actually present, the correction to SdHO would have been up to $> 3$ ($> 5$) times larger than observed in sample \rom{1} (sample \rom{2}). 
We thus conclude that the proposal of \rref{mani:2007} is irrelevant to our findings.

In summary, we have studied the photoresistance in high-quality 2DEG subject to low temperatures and high microwave frequencies, which allowed us to overlap MIRO and SdHO over multiple harmonics of the cyclotron resonance.
Our data revealed pronounced modulation of the SdHO which is periodic in $\eac$, with the period equal to unity, and the phase matching that of MIRO.
This result does not fit existing theories considering either magnetoabsorption or photoresistance. 
Most remarkably, we have found that once the exponential factors are eliminated, the oscillatory part of the SdHO amplitude matches that of MIRO \emph{quantitatively}, without any adjustable parameters. 
This finding allowed us to deduce an empirical relation for the SdHO amplitude in irradiated 2DEG, given by \reqs{eq.s}{eq.alpha}.
Taken together, our study reveals that current understanding of SdHO in irradiated 2DEG is lacking, calling for further investigations.

\begin{acknowledgments}
We thank G. Jones, S. Hannas, T. Murphy, J. Park, and D. Smirnov for technical assistance with experiments.
The work at Minnesota (Purdue) was supported by the U.S. Department of Energy, Office of Science, Basic Energy Sciences, under Award \# ER 46640-SC0002567 (DE-SC0006671).
The work at Princeton was partially funded by the Gordon and Betty Moore Foundation and by the NSF MRSEC Program through the Princeton Center for Complex Materials (DMR-0819860).
A portion of this work was performed at the National High Magnetic Field Laboratory (NHMFL), which is supported by NSF Cooperative Agreement No. DMR-0654118, by the State of Florida, and by the DOE.
Q.S. acknowledges Allen M. Goldman fellowship.
\end{acknowledgments}

%\bibliographystyle{../../apsrev}
%\bibliography{../../bibRMP1qs,footnotes}

\begin{thebibliography}{67}
\bibitem[{not({\natexlab{a}})}]{note:100}
\bibinfo{note}{$\pc = \pc_+ + \pc_-,~\pc_{\pm} = e^2\Eac^2v_F^2/2\epseff
  \hbar^2 \omega^4[(1 \pm \eac^{-1})^2+(\omega\tem)^{-2}]$, where
  $\tem^{-1}=n_e e^2/2\sqrt{\epseff}\epsilon_0m^\star c$,
  $2\sqrt{\epseff}=\sqrt{\varepsilon}+1$, $\varepsilon=12.8$ is the dielectric
  constant of GaAs, $v_F$ is the Fermi velocity, and $\Eac$ is the microwave
  electric field.}
\bibitem[{not({\natexlab{b}})}]{note:7}
\bibinfo{note}{The rate of scattering on angle $\theta$ can be expressed in
  terms of angular harmonics, $\tau_n=\tau_{-n}$, as $\tau_\theta^{-1} =
  \sum\tau_n^{-1}e^{i n\theta}$.}
\bibitem[{not({\natexlab{d}})}]{note:1}
\bibinfo{note}{This contribution, which directly affects SdHO, was not
  considered in relation to MIRO, originating from corrections of order
  $\o(\lambda^2)$.}
\bibitem[{not({\natexlab{e}})}]{note:202}
\bibinfo{note}{Under conditions of our experiment, in sample \rom{1} (\rom{2})
  we estimate $\tq$ to be 2 (3) ps from SdHO and 4.4 (8.2) ps from MIRO.
  However, accurate determination of SdHO $\tq$ requires extremely slow sweeps
  at very low $B$ which were not attempted because the exact knowledge of $\tq$
  is not crucial for our study.}
\bibitem[{not({\natexlab{g}})}]{note:5}
\bibinfo{note}{Strictly speaking, $\rhosmo \neq \rhosm$ because of
  radiation-induced heating and $T$-dependence of $\rhosm$.}
\bibitem[{not({\natexlab{h}})}]{note:301}
\bibinfo{note}{While all samples surveyed have shown correlation between MIRO
  and SdHO, in many of them MIRO rapidly decayed with increasing $f$. As a
  result, modulation of the SdHO was observed only up to $\eac \approx 2$,
  where the excess heating due to the proximity to the cyclotron resonance
  precluded quantitative analysis of the waveforms.}
\bibitem[{not({\natexlab{i}})}]{note:101}
\bibinfo{note}{We are not suggesting that this scenario is justified.}
\end{thebibliography}

\end{document}